\documentclass[aip,rsi,reprint,graphicx]{revtex4-1}
\usepackage{graphicx}
\usepackage{upgreek}

\begin{document}
\title{Optically transparent solid electrodes for precision Penning traps}
\author{M. Wiesel}
\affiliation{GSI Helmholtzzentrum f\"ur Schwerionenforschung, 64291 Darmstadt, Germany}
\affiliation{Physikalisches Institut, Ruprecht Karls-Universit\"at, 69120 Heidelberg, Germany}
\affiliation{Institut f\"ur Angewandte Physik, TU Darmstadt, 64291 Darmstadt, Germany}
\author{G. Birkl}
\affiliation{Institut f\"ur Angewandte Physik, TU Darmstadt, 64291 Darmstadt, Germany}
\author{M.S. Ebrahimi}
\affiliation{GSI Helmholtzzentrum f\"ur Schwerionenforschung, 64291 Darmstadt, Germany}
\affiliation{Physikalisches Institut, Ruprecht Karls-Universit\"at, 69120 Heidelberg, Germany}
\author{A. Martin}
\affiliation{Institut f\"ur Angewandte Physik, TU Darmstadt, 64291 Darmstadt, Germany}
\author{W. Quint}
\affiliation{GSI Helmholtzzentrum f\"ur Schwerionenforschung, 64291 Darmstadt, Germany}
\affiliation{Physikalisches Institut, Ruprecht Karls-Universit\"at, 69120 Heidelberg, Germany}
\author{N. Stallkamp}
\affiliation{GSI Helmholtzzentrum f\"ur Schwerionenforschung, 64291 Darmstadt, Germany}
\affiliation{Helmholtz-Institut Jena, 07743 Jena, Germany}
\author{M. Vogel}
\affiliation{GSI Helmholtzzentrum f\"ur Schwerionenforschung, 64291 Darmstadt, Germany}
\affiliation{Helmholtz-Institut Jena, 07743 Jena, Germany}

\date{\today}

\begin{abstract}
We have conceived, built, and operated a cryogenic Penning trap with an electrically conducting yet optically transparent solid electrode. The trap, dedicated to spectroscopy and imaging of confined particles under large solid angles is of 'half-open' design with one open endcap and one closed endcap that mainly consists of a glass window coated with a highly transparent conductive layer. This arrangement allows for trapping of externally or internally produced particles, yields flexible access for optical excitation and efficient light collection from the trapping region. At the same time, it is electrically closed and ensures long-term ion confinement under well-defined conditions. With its superior surface quality and its high as well as homogeneous optical transmission, the window electrode is an excellent replacement for partially transmissive electrodes that use holes, slits, metallic meshes and the like.
\end{abstract}

\pacs{37.10.Ty Ion trapping }
\maketitle

\section{Introduction}
Penning traps are valuable tools for precision spectroscopy of confined particles, largely owing to the multitude of available techniques for ion confinement and cooling \cite{werth,gho}. Confinement of internally or externally produced ions leads to a localisation of the species of interest in a small volume of space for long times. Cooling of these ions reduces shifts and transition line broadenings due to the Doppler effect \cite{dem}, giving access to spectroscopy with high resolution and precision. 
To this end, a multitude of cooling techniques have been developed \cite{ita}. The most common techniques used in Penning traps are resistive cooling \cite{res,vog} and several forms of laser cooling \cite{ic,ita2,buch}, which either require conducting surfaces close to the ions, good optical access, or both. 

\begin{figure}[h!]
\begin{center}
\centering
\includegraphics[width=0.65\columnwidth]{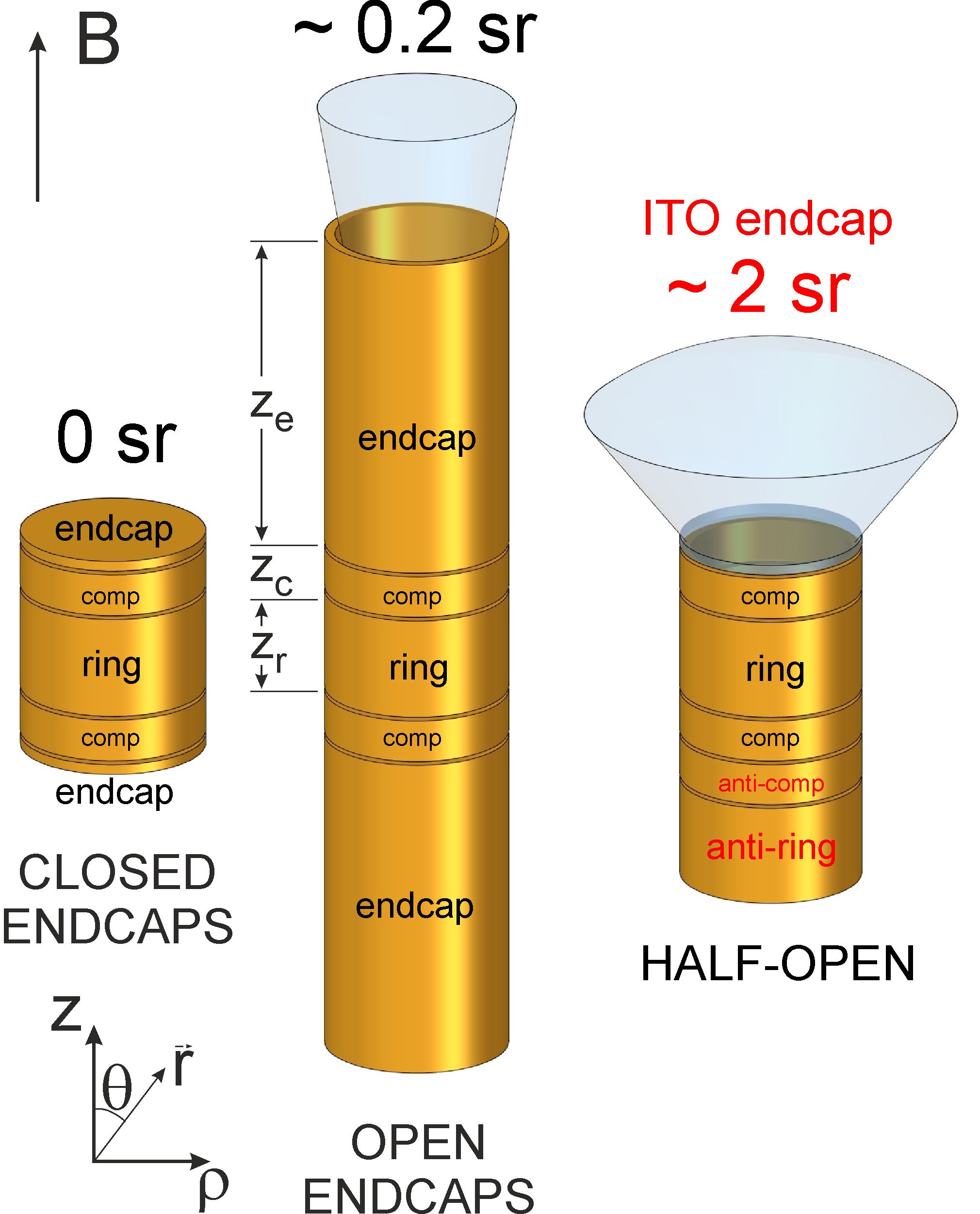}
\caption{Comparison of the designs of cylindrical Penning traps with closed endcaps, open endcaps and the half-open design, with the typical solid angles of optical access indicated.}
\label{one}
\end{center}
\end{figure}

The optimum design of a Penning trap for spectroscopy is commonly a compromise between two opposing ideals, namely electrically closed while optically open. Electrically closed in this context means that voltages applied to the trap electrodes create a well-defined potential well which confines the ions for long durations. Often the ideal is an electrostatic field of quadrupolar shape, such that the axial confinement is harmonic. Hyperbolic Penning traps have this property by design, but are almost completely closed mechanically and optically (similar to the closed-endcap cylindrical Penning trap in figure \ref{one}, left), which prevents both loading of externally produced ions and optical access \cite{gho}. One common solution is to introduce artificial openings such as holes in the electrodes. These however adversely affect the confining potential and hence need to be kept small. A cylindrical Penning trap with open endcaps \cite{gab89} (figure \ref{one}, centre) solves this problem to some extent while retaining well-defined confinement conditions close to the trap centre. However, the light collection efficiency is limited by the small ratio of diameter to length of the arrangement. Radial openings as used in experiments as discussed in ref. \cite{bhar,spec,spec2} circumvent this limitation to some extent, but are unsuitable when high precision of the confining fields is required. Also, they necessitate either complicated light collection schemes \cite{bhar} or radial access to the trap \cite{spec,spec2}.
 Optical access to confined particles has been part of the motivation for the development of planar Penning traps \cite{pla,pla2,gal,bush,gabbash} as well as of numerous other types of traps such as wire Penning traps \cite{cast}, stylus traps \cite{wine} and others \cite{a1,a2,a3}. In such a configuration, light collection can be highly efficient and approach solid angles of $4\pi$ sr. However, such traps unfortunately are unsuited for precision spectroscopy of large ensembles of ions as for example discussed in refs. \cite{pra1,pra2}.

We have conceived, built, and operated a so-called 'half-open' cylindrical Penning trap \cite{davv} (figure \ref{one}, right) with the favourable electrical properties of a compensated open-endcap Penning trap, while at the same time surpassing its solid angle of optical access by more than an order of magnitude. While in the original configuration, the endcap has been electrically closed by a metal mesh, in the advanced design presented here, it features a window with transparent yet conductive coating, which has been characterized in a series of measurements, particularly with regard to its use in cryogenic Penning trap experiments.    

\section{Setup}
The present studies have been performed at the ARTEMIS setup \cite{pra1,pra2} located at the HITRAP facility \cite{hitrap} at GSI / FAIR (Darmstadt, Germany). The Penning trap is located in the field centre of a superconducting magnet and cooled to liquid-helium temperature. The magnet provides a static homogeneous magnetic field of 7\,T which assures radial confinement of charged particles in the Penning trap on account of the Lorentz force. Axial confinement is provided by the electrostatic field inside the trap which is created by voltages applied to trap electrodes, chosen to create a harmonic potential well \cite{gab89}.
We have detailed the concept of a half-open cylindrical Penning trap previously \cite{davv}. Briefly, it is a variation of a closed cylindrical Penning trap in which one closed endcap is replaced by open cylindrical electrodes ('anti-comp' and 'anti-ring' in figure \ref{one}, right) to allow injection and ejection of particles, while the other endcap is electrically closed but optically transparent, which in the present case is realized by a window with conductive coating. This dedicated arrangement of electrodes brings the trap centre and hence the confinement region in close proximity to the optically open endcap, and thus leads to a largely enhanced solid angle of detection as compared to the standard open-endcap design (see figure \ref{one}, centre). 

To quantify this statement, we look at the typical dimensions of such traps. Let the inner trap radius be $\rho_0$ and the lengths of the ring, compensation and endcap electrodes be $z_r$, $z_c$ and $z_e$, respectively. Then, ignoring the small gaps between electrodes, the distance $z$ of the trap centre from the end surface of the trap endcap is given by {$z=z_r/2 + z_c + z_e$} (see figure \ref{one}). In the 'half-open' trap with $z_e=0$, this distance is about $z \approx 1.5\rho_0$, while in standard open-endcap cylindrical traps the endcaps need to be very long to achieve comparable harmonicity ($z_e$ at least about $4z_0$ while $z_0 \approx \rho_0$, see ref. \cite{gab89}), such that this distance is about $z \approx 5\rho_0$. The full opening angle $\Theta$ of the light cone from the trap center is given by 
\begin{equation}
\Theta = 2\tan^{-1}\left(\frac{\rho_0}{z}\right),
\end{equation}
and the corresponding solid angle of detection $\Omega$ is given by 
\begin{equation}
\Omega = 4 \pi \sin^2 \left( \frac{\Theta}{4}  \right) .
\end{equation}
Hence, the ratio of opening angles $\Theta$ between the traps is about 10/3 and the corresponding ratio of solid angles is about 11 in favour of the half-open design. 

The advantage of light detection along the trap axis as possible in an optically open endcap trap as compared to radial light detection is even more pronounced when considering the angular distribution of magnetic dipole radiation emitted from confined ions, as foreseen in the present experiment. In this case, the fluorescence light of interest emitted from the trap centre has an angular distribution given by \cite{jack}
\begin{equation}
I(\theta){\rm d}\theta \propto (1+\cos^2\theta) {\rm d}\theta, \label{eq:distr}
\end{equation}
where $\theta$ is the spherical polar angle with respect to the central trap axis. In the absence of collisions, this directional characteristics enhances emission along the axis of the magnetic field (i.e. the trap axis) by a factor of 2 with respect to radial emission \cite{jack}, which further increases the ratio of light collected through the window to the total light emitted.

To retain the electrostatic trapping potential of the trap, the inner surface of the window is coated with a thin conductive layer of indium tin oxide (ITO). 
\begin{figure}[h!]
\begin{center}
\centering
\includegraphics[width=\columnwidth]{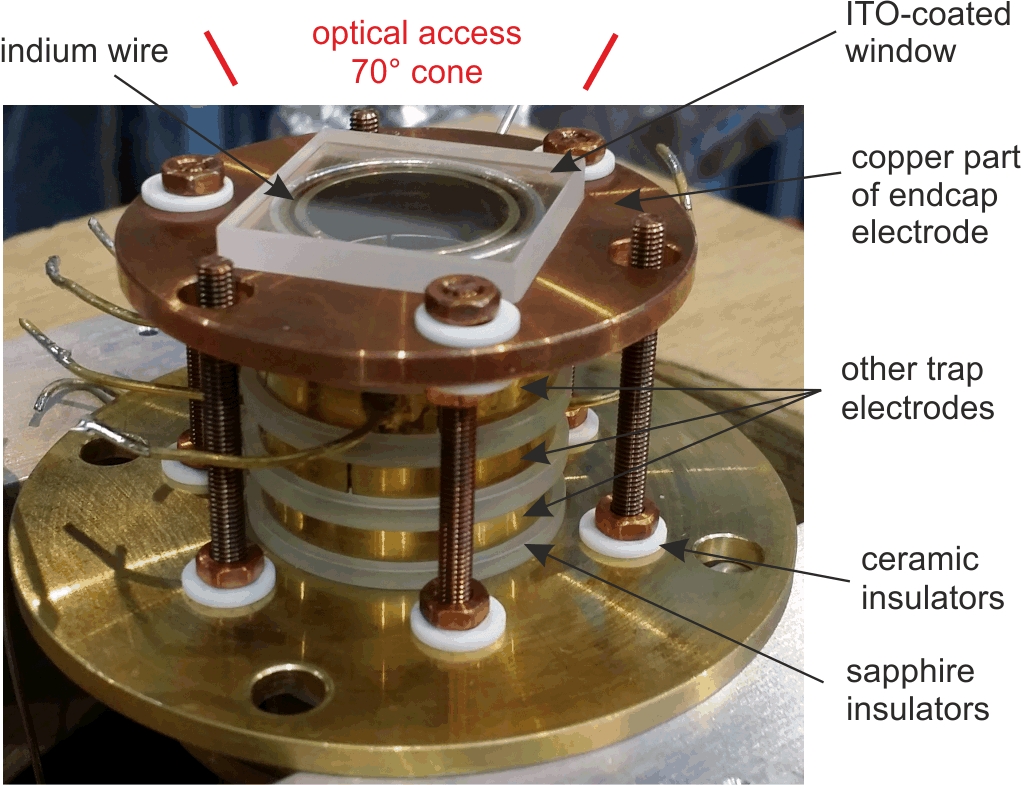}
\caption{Picture of the relevant part of the Penning trap with the closed endcap consisting of a copper part and the coated window giving 70$^{o}$ optical access to the trap centre. The linear dimension of the glass substrate is $25.4$\,mm}
\label{oneb}
\end{center}
\end{figure}
ITO is a doped n-type semiconductor with a bandgap of around 4\,eV and is hence largely transparent in the visible part of the spectrum. It typically consists of about 75\,\% indium (In), 17\,\% oxygen (O$_2$) and 8\,\% tin (Sn) coated on a substrate. Several coating techniques are in use, including sputtering, spray coating and vapour deposition, all of which allow to fully coat or to produce 2D structures on the substrate surface, including bent substrates. Besides ITO (In$_2$O$_3$:Sn), a number of other coatings are available, mainly FTO (SnO$_2$:F) and ATO (SnO$_2$:Sb) which have roughly similar properties, for details see \cite{itop}. ITO has a bulk resistance of about $10^{-5}$\,$\Omega$m (three orders of magnitude higher than copper), which in the present case corresponds to a sheet resistance of around 80\,$\Omega$/sq at a layer thickness of 125\,nm \cite{itop}.  
\begin{figure}[h!]
\begin{center}
\centering
\includegraphics[width=0.9\columnwidth]{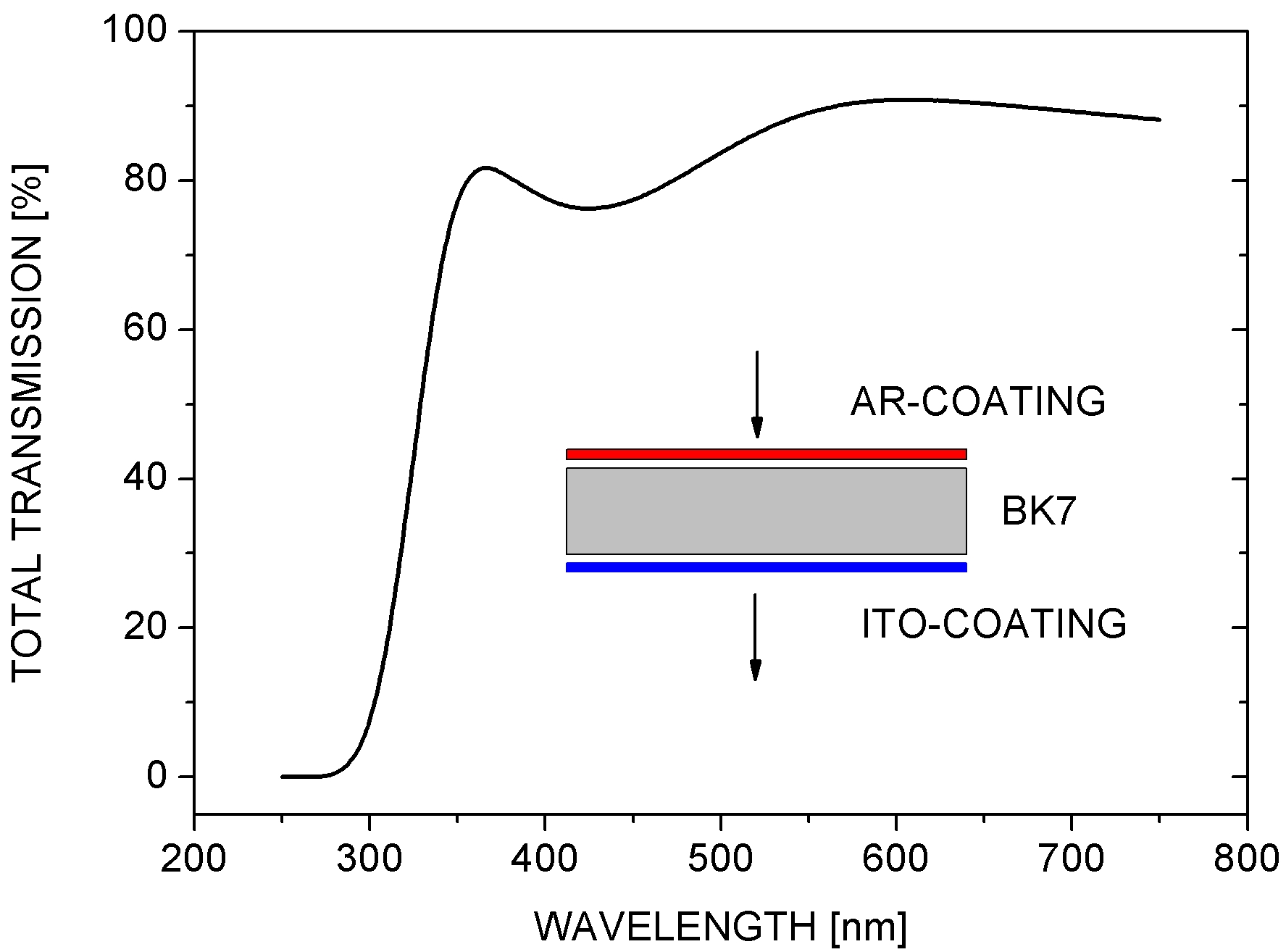}
\caption{Total transmission of the ITO-coated window as a function of wavelength, according to the manufacturer \cite{thor}. The inset shows the window coating structure.}
\label{two}
\end{center}
\end{figure}
The present window is a flat, fully ITO-coated square conductive window (ThorLabs WTSQ11050-A). It has dimensions $25.4 \times 25.4 \times 4$\,mm and consists of an N-BK 7 substrate with an anti-reflection coating on one and an ITO coating on the other surface (see the inset in figure \ref{two}). Note, that in situations where high thermal conductivity is required, a sapphire substrate can be used. The optical transmission of the coating is comparable to or better than that of typical metallic meshes used in similar cases. Importantly, the conducting surface is geometrically better defined by several orders of magnitude (presently at a surface flatness of 80\,nm or better), allowing highly harmonic confining potentials. This is an important feature since in our trap design, the ions are confined in close proximity to the closed endcap such that the potential at their position is sensitive to imperfections of the electrode's conducting surface. 
The potentially high grade of definition of glas surfaces also facilitates applications in small traps where confined particles are necessarily close to coated glas surfaces.  
The electrical contact between that surface and the metal part of the endcap is made with an indium wire pressed between the two flat surfaces, see figure \ref{oneb}. The window is fastened to the copper part of the endcap electrode by pressing down a second copper ring placed on top of the window, making the indium contact.

\section{Conducting Window Characterization}
Figure \ref{two} shows the total transmission of the ITO-coated window (including the antireflection coating on the backside) as a function of wavelength based on data provided by the manufacturer \cite{thor}. 
For the experiments envisaged with the present trap, a window with anti-reflection coating optimized for the visible wavelength region has been chosen. The transmission of the window surpasses that of typical metal meshes (50\%-70\%) for wavelengths above 350\,nm.

To characterize the performance of the window as an electrode in a cryogenic Penning trap, several measurements have been performed. When introducing components in this environment, it is often unknown whether they will withstand the conditions of cryogenic temperature at high magnetic field and operate properly. 
\begin{figure}[h!]
\begin{center}
\centering
\includegraphics[width=0.9\columnwidth]{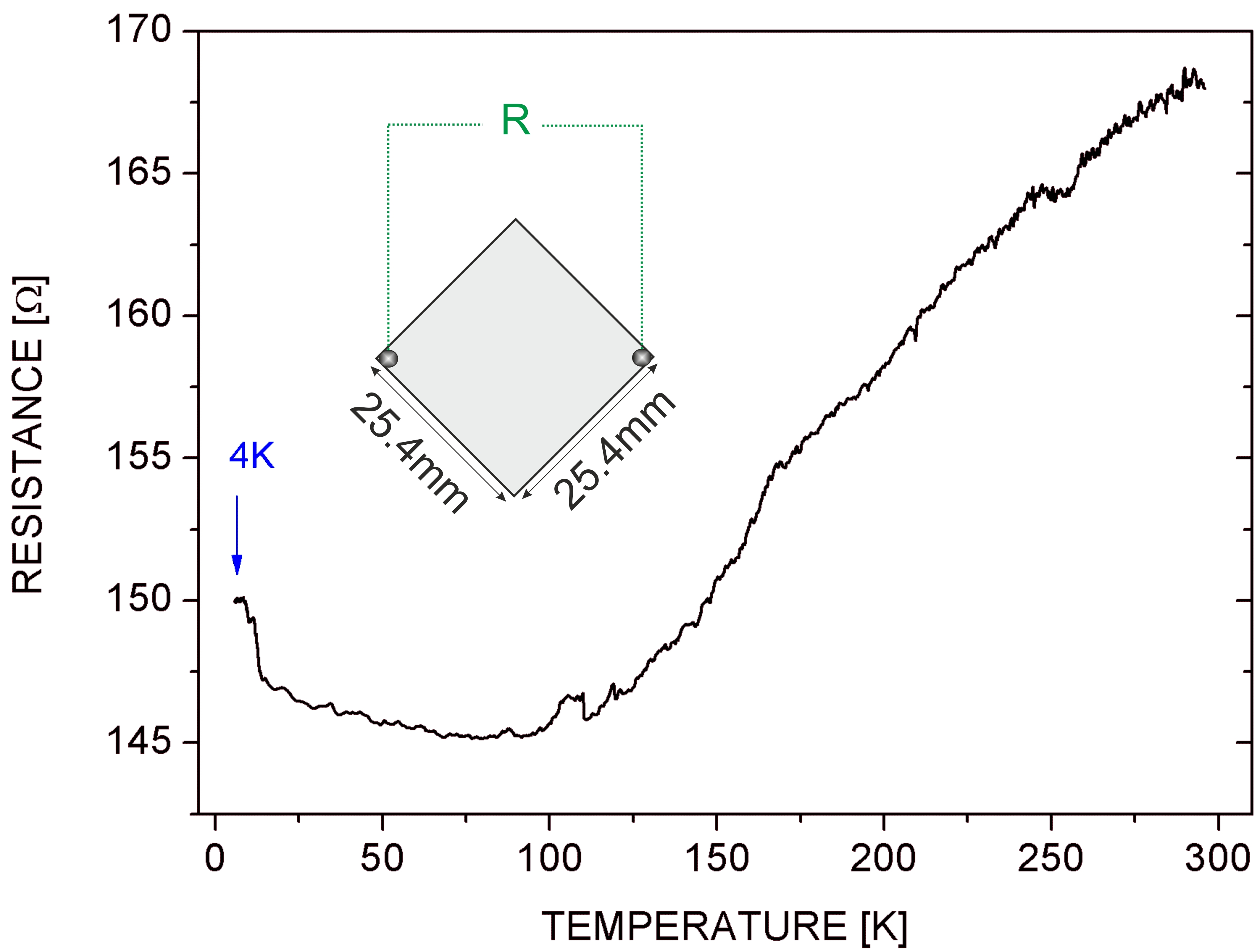}
\caption{Measured resistance across the ITO-coated trap electrode as a function of temperature. The inset shows the geometry of the measurement.}
\label{three}
\end{center}
\end{figure}
Specifically, cooling down and warming up between 300\,K and 4\,K may introduce significant stress. It is also \textit{a priori} unclear whether issues such as charge build-up of the window arise, and whether cryogenic vacua might be spoiled.
For our specific application, we were also interested in the possibility to use the ITO-coated electrode as a Faraday cup for ion detection, which requires it to withstand bombardment of ions. Since the coating thickness is of the order of 100\,nm only, it is important to test whether highly charged ions at keV kinetic energies will sputter or otherwise damage the surface. Details on these tests are given in section \ref{secdetect}

Figure \ref{three} shows the measured resistance of the ITO-coated trap electrode as a function of temperature between 300\,K and 4\,K. The resistance was measured between two diagonally opposite corners of the coating, as shown in the inset of figure \ref{three}. 
Overall, the measured values fit the expectance when the coating is treated as a conductor of length $L=\sqrt{2} l$ (where $l=25.4$\,mm is the linear dimension of the window and $L$ is the distance between the contact points) with an effective cross section of $A=L/2 \times h$, in which $h \approx 125$\,nm is the layer thickness. Using $R=\rho L / A$ with the ITO bulk resistance at room temperature of $\rho \approx 10^{-5}$\,$\Omega$m yields $R \approx 160$\,$\Omega$, close to the measured value.

Obviously, the coating withstands cooling down to 4\,K and displays a resistance which is reduced by around 10\% when comparing 300\,K to 4\,K. Above 125\,K, the resistance-temperature dependence is close to linear with a temperature coefficient of 0.1298(5)\,$\Omega$/K which is a relative value of $8.125 \times 10^{-4}$. The overall shape of the measured resistance curve follows the Bloch-Gr\"uneisen law, as expected for metallic layers like the present one \cite{lilin,chinesen}. 
Note that for much smaller layer thickness below 40\,nm, significantly different behaviour has been reported \cite{china}. 
The observed increase of resistance below about 100\,K is also in accordance with measurements of similar ITO films \cite{lilin,chinesen} and is attributed to effects of weak localisation and electron-electron interaction \cite{lilin}. The temperature dependence of the resistance is expected to be different when an external magnetic field is applied, this however is a minor effect \cite{chinesen}.
Measurements of the electrical conductivity have been reported \cite{ieee} up to a frequency of 20\,GHz and show that it remains roughly at the dc value of about $0.5 \times 10^6$\,S/m. The skin depth of ITO up to 20\,GHz is above 1\,$\mu$m, and hence much larger than the layer thickness \cite{ieee}. From this, the ITO layer can be expected to behave similarly throughout the whole radio-frequency domain.

\section{Ion Detection: The conducting Window as a Faraday Cup}
\label{secdetect}
Since trap electrodes are often used for destructive or non-destructive detection of charged particles (for a recent example see e.g. \cite{nondestructive}) as well, we have investigated the behaviour of the coating when used as a Faraday cup. 
\begin{figure}[h!]
\begin{center}
\centering
\includegraphics[width=0.9\columnwidth]{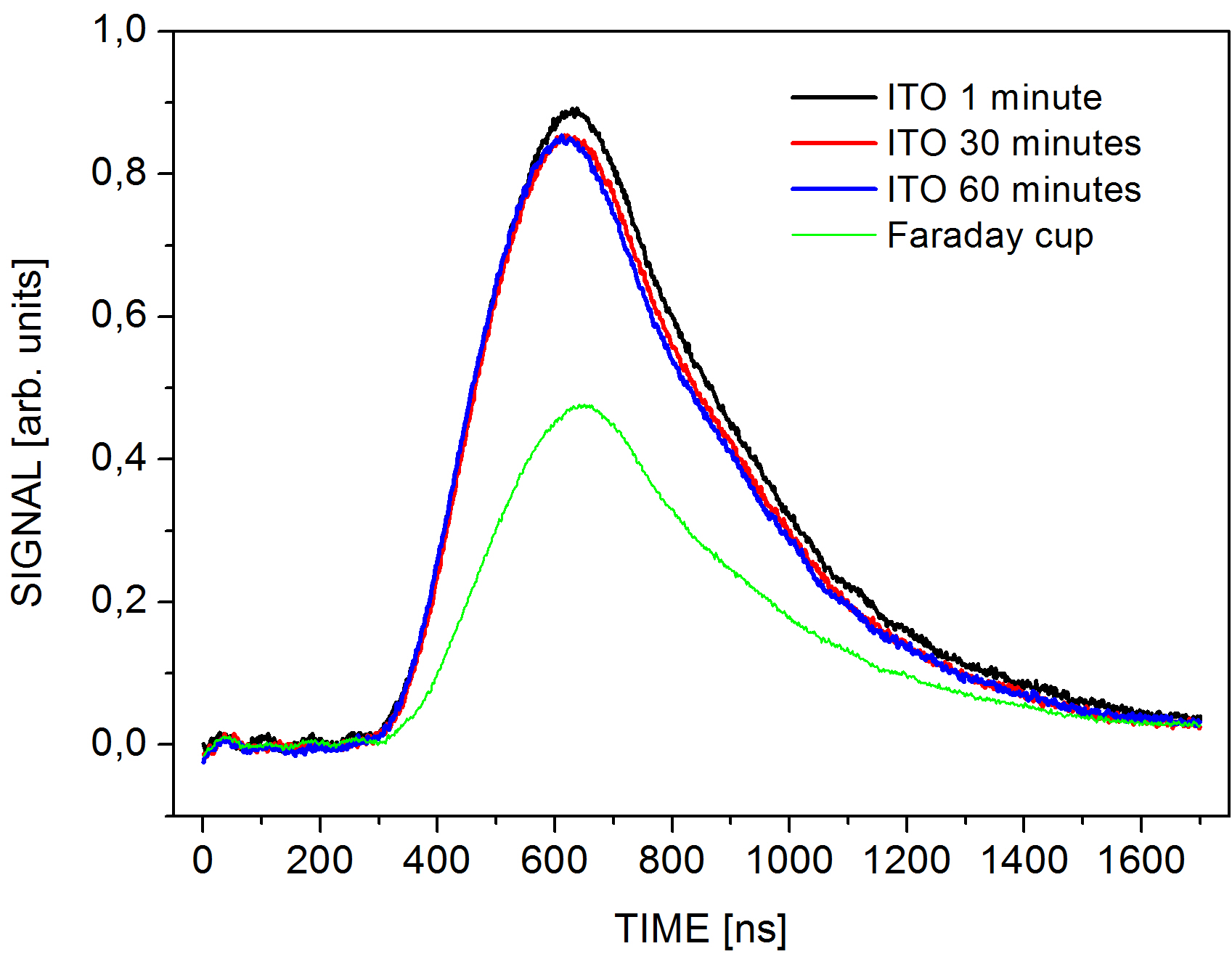}
\caption{Ion detection with the ITO coated window operated as a Faraday cup: Temporal structure of ion signals after three different periods of ion bombardment. For comparison, the signal of a metal cup electrode is shown. The difference in signal height between ITO and Faraday cup is due to different overall resistances of the arrangement.}
\label{four}
\end{center}
\end{figure}
In order to make a stringent statement about the resilience and durability of the coating under impinging particles, we have used a commercial electron beam ion source (EBIS) at the HILITE cryogenic Penning trap setup \cite{hilite} to expose the surface to highly charged ions of considerable energy. In this measurement, we have used highly charged ions of charge states up to around 16+ with an ion current of several nA at a kinetic energy of 5\,keV per charge. The ion charge current across the conductive layer of the window is translated into a voltage signal by a Femto DHPCA-100 current amplifier and read out on an oscilloscope. 
Figure \ref{four} shows the corresponding time-resolved signals of the impinging ion bunches. Three measurements for the ITO-coated window after one, thirty and sixty minutes of ion bombardment are shown. For comparison, an additional measurement was performed with a Faraday cup in its commercial configuration (green line in figure \ref{four}).  All ITO-window signals are very similar in shape and show no significant degradation over time, even under the harsh circumstances given. The metal electrode signal is lower in amplitude due to a different overall resistance of the measurement arrangement, but shows an almost identical temporal structure. Hence, the ITO-coated window can be used for detection of charged particles in cryogenic trap experiments with a behaviour identical to a regular Faraday cup.

\section{Confinement of Ions in a half-open trap with ITO-coated window endcap}
In order to demonstrate the operation of a Penning trap with an ITO-coated window as endcap, we trapped highly charged argon ions over durations of hours in the ARTEMIS setup \cite{pra1,pra2}. 
\begin{figure}[h!]
\begin{center}
\centering
\includegraphics[width=\columnwidth]{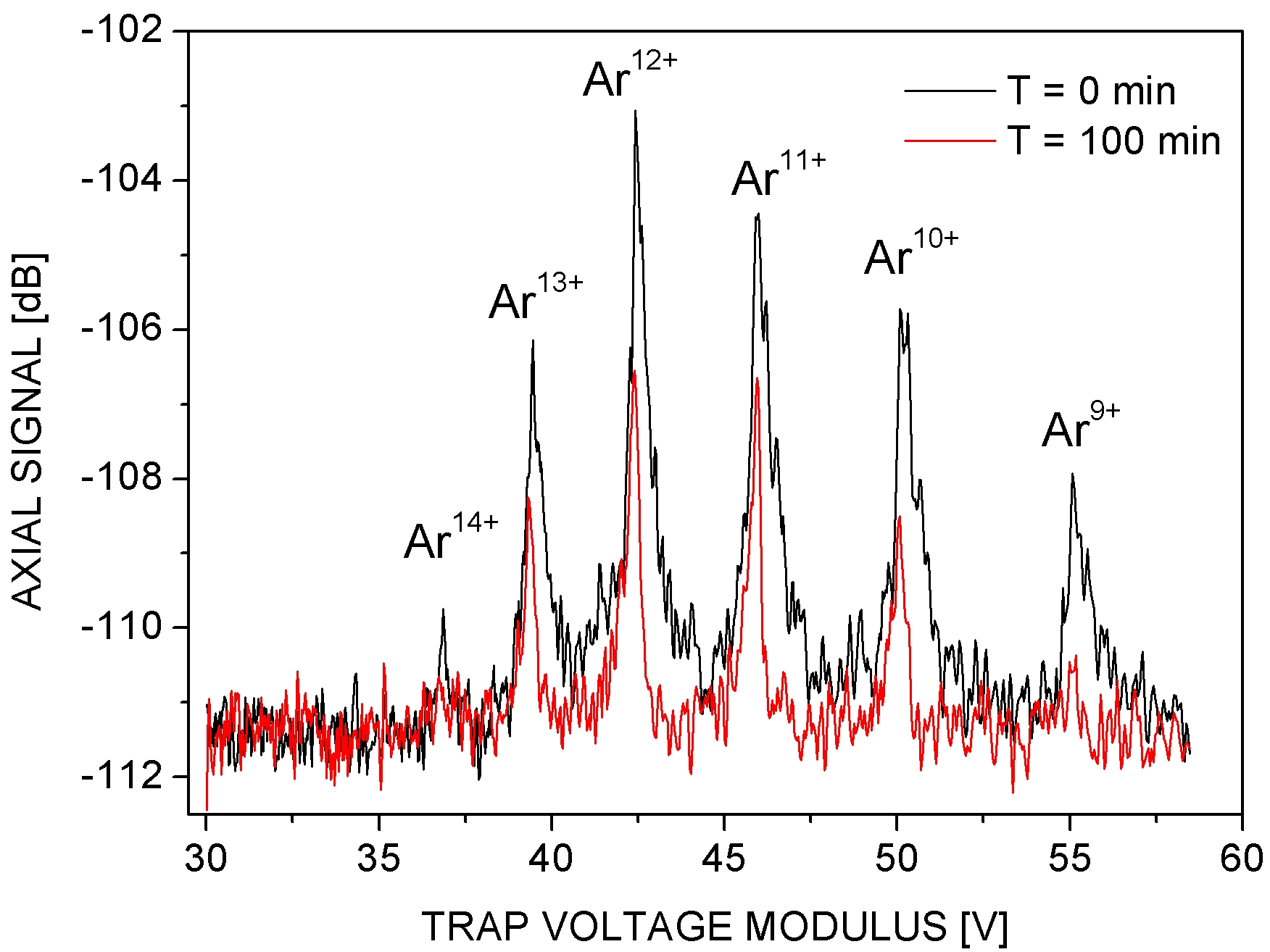}
\caption{Charge state spectra of argon ions confined in the half-open Penning trap with the ITO-coated window endcap after two different confinement times.}
\label{five}
\end{center}
\end{figure}
We generated the ions in a series of charge states via charge breeding in the creation trap of ARTEMIS (open endcap trap) and then transferred the ions to the spectroscopy trap of ARTEMIS (half-open trap with ITO-coated window). The creation part of the trap setup is directly adjacent to the half-open trap used for spectroscopy and shares the same vacuum.

Figure \ref{five} shows two spectra of highly charged argon ions confined in the half-open Penning trap with the ITO-coated window, as taken by a resonant-circuit detection scheme which non-destructively detects the current induced in trap electrodes due to the oscillation of the confined ions. Stable confinement was possible for extended periods of time, as apparent from the two spectra taken 100 minutes apart. 
The reduction of the signal height and slight shift of spectral position after 100 minutes is a well-known effect of resistive cooling by the resonant detection circuit when taking subsequent spectra \cite{vog} and does not reflect ion loss. 

Also, the stable confinement over long durations indicates that there is no significant charge build-up on the window (as also expected from its sufficiently high conductance) since this would change the spectral appearance of the stored ions.
In a different series of measurements, we have observed charge state lifetimes of weeks, from which the residual gas pressure in the traps was estimated to be on the $10^{-16}$\,mbar level, leading to the conclusion that the ITO-coated window at cryogenic temperatures does not spoil the vacuum on that level. From the measured dc resistance, the observed signals when used as a Faraday cup, and from the stable confinement over long periods of time, we can exclude charging up of the window, which is also expected to be true when laser light passes through the window. From our observations with visible laser light on the mW scale of power, there is no obvious degradation of the coating.

\section{Summary and Conclusion}
With this work we have introduced a novel realisation of Penning traps for precision spectroscopy. We have conceived, built, and operated a half-open cylindrical Penning trap in which one endcap is made of a window with an electrically conductive yet optically transparent coating. This allows for a stable confinement of charged particles under well-defined conditions and efficient optical access to the confined particles under a large solid angle. This arrangement lends itself towards demanding applications such as optical precision spectroscopy of confined ions requiring efficient light collection. We have shown stable confinement of highly charged ions in such a trap and have studied the resilience and durability of an ITO-coated window used as an electrode in a cryogenic Penning trap environment. The use of such conductive coatings will not remain limited to single plane endcaps, however, such that a large variety of applications appears possible. This is true in particular when coatings with predefined geometrical patterns are considered, e.g. for planar trapping structures on transparent substrates, multiple realisations of traps on a single substrate, or highly integrated systems, combining electromagnetic and optical sub-systems. From our observations, we see no fundamental obstacles to a Penning trap that is completely manufactured from coated substrates, provided that the surfaces can be properly contacted.

\begin{acknowledgments}
We gratefully acknowledge the work of all members of the ARTEMIS collaboration. We also acknowledge financial support by the DFG under contract BI 647/5-1 and by HIC for FAIR within the LOEWE program of the federal state Hessen. A.M. and M.W. acknowledge support from HGS-HIRe. N.S. acknowledges support from RS-APS Jena.
The experiments have been performed within the framework of the HITRAP facility at GSI and FAIR, Darmstadt.

\end{acknowledgments}

\end{document}